\documentclass[aip,reprint,superscriptaddress,noeprint,showkeys]{revtex4-2}

\usepackage[hidelinks]{hyperref}% add hypertext capabilities
\usepackage[dvipsnames]{xcolor}
\hypersetup{
    colorlinks,
    linkcolor={blue},
    citecolor={blue},
    urlcolor={blue},
}
\usepackage{graphicx}
\usepackage{amsmath,braket,graphicx,mathtools,amssymb}
\usepackage{amsfonts}
\usepackage{braket}
\usepackage{bm}

\newcommand{\CrPS}{$\mathrm{CrPS_4}$}

\newcommand{\red}[1]{\textcolor{Black}{#1}}

\newcommand{\mtwo}{$\downarrow\uparrow$}
\newcommand{\mthree}{$\downarrow\uparrow\downarrow$}
\newcommand{\mfour}{$\downarrow\uparrow\downarrow\uparrow$}

\newcommand{\msix}{$\downarrow\uparrow\downarrow\uparrow\downarrow\uparrow$}

\newcommand{\ptwo}{$\uparrow\downarrow$}
\newcommand{\pthree}{$\uparrow\downarrow\uparrow$}
\newcommand{\pfour}{$\uparrow\downarrow\uparrow\downarrow$}

\newcommand{\psix}{$\uparrow\downarrow\uparrow\downarrow\uparrow\downarrow$}

\usepackage{multibib}
\newcites{Methods}{Methods References}
\begin{document}

%\preprint{}

%\title{Adjustable lateral exchange bias by control over atomically thin\\antiferromagnetic domains}
%\title{Imaging and control of antiphase domains through layer parity exchange in 2D anti}
\title{Configurable antiferromagnetic domains and lateral exchange bias\\ in atomically thin CrPS$_4$}

\author{Yu-Xuan Wang}
%\email{yuxuan@bc.edu}
\author{Thomas K. M. Graham}%grahamtb@bc.edu
\affiliation{Department of Physics, Boston College, Chestnut Hill, MA 02467, USA}

\author{Ricardo Rama-Eiroa}
\affiliation{Donostia International Physics Center (DIPC), 20018, Donostia-San Sebastián, Basque Country, Spain}
\affiliation{Institute for Condensed Matter and Complex Systems, School of Physics and Astronomy, University of
Edinburgh, Edinburgh EH9 3FD, United Kingdom}

\author{Md Ariful Islam}
\affiliation{Department of Physics, Applied Physics and Astronomy, Binghamton University, Binghamton, NY 13902, USA}

\author{Mohammad H. Badarneh}
\affiliation{Institute for Condensed Matter and Complex Systems, School of Physics and Astronomy, University of
Edinburgh, Edinburgh EH9 3FD, United Kingdom}

\author{Rafael Nunes Gontijo}
\author{Ganesh Prasad Tiwari}
\author{Tibendra Adhikari}
\affiliation{Department of Physics, Applied Physics and Astronomy, Binghamton University, Binghamton, NY 13902, USA}

\author{Xin-Yue Zhang}
\affiliation{Department of Physics, Boston College, Chestnut Hill, MA 02467, USA}

\author{Kenji Watanabe}%WATANABE.Kenji.AML@nims.go.jp
\affiliation{Research Center for Electronic and Optical Materials, National Institute for Materials Science, 1-1 Namiki, Tsukuba 305-0044, Japan}

\author{Takashi Taniguchi}%TANIGUCHI.Takashi@nims.go.jp
\affiliation{Research Center for Materials Nanoarchitectonics, National Institute for Materials Science,  1-1 Namiki, Tsukuba 305-0044, Japan}

\author{Claire Besson}
\affiliation{Department of Chemistry, Binghamton University,  Binghamton, NY 13902, USA}

\author{Elton J. G. Santos}
\affiliation{Institute for Condensed Matter and Complex Systems, School of Physics and Astronomy, University of
Edinburgh, Edinburgh EH9 3FD, United Kingdom}
\affiliation{Higgs Centre for Theoretical Physics, The University of Edinburgh, Edinburgh EH9 3FD, UK}
\affiliation{Donostia International Physics Center (DIPC), 20018, Donostia-San Sebastián, Basque Country, Spain}

\author{Zhong Lin}
\affiliation{Department of Physics, Applied Physics and Astronomy, Binghamton University, Binghamton, NY 13902, USA}
\affiliation{Materials Science and Engineering Program, Binghamton University,  Binghamton, NY 13902, USA}

\author{Brian B. Zhou}
\email{brian.zhou@bc.edu}
\affiliation{Department of Physics, Boston College, Chestnut Hill, MA 02467, USA}

\date{\today}
\begin{abstract}
%insert abstract here

Interfacial exchange coupling between antiferromagnets (AFMs) and ferromagnets (FMs) crucially makes it possible to shift the FM hysteresis, known as exchange bias, and to switch AFM states. Two-dimensional magnets unlock opportunities to combine AFM and FM materials; however, the buried AFM–FM interfaces obtained by stacking remains challenging to understand. Here we demonstrate interfacial control via intralayer exchange coupling in the layered AFM CrPS$_4$, where connected even and odd layers realize pristine lateral interfaces between AFM-like and FM-like regions. We distinguish antiphase even-layer states by scanning nitrogen-vacancy centre (NV) magnetometry due to a weak surface magnetization. This surface magnetization enables control over the even-layer state, with different regions switching at distinct fields due to their own lateral couplings. We toggle three AFM domains adjacent to a FM-like region and demonstrate a tunable multilevel exchange bias. Our nanoscale visualization unveils the microscopic origins of exchange bias and advances single two-dimensional crystals for hybrid AFM–FM technologies.

\end{abstract}

%odd layers is tunable through the configuration of the surrounding even-layer AFM domains, which can be toggled on-demand by . 

%\pacs{}
%\keywords{Quantum sensing, current imaging, nitrogen-vacancy center, two-dimensional semimetal}

\maketitle
%\section{}
%\begin{center}
%\rule{250pt}{1pt}
%\end{center}
%\newpage
%with avenues for tunability through defect and interfacial engineering

% pn junction
% thermoelectric

Interfaces between materials with diverging properties can engender novel functionalities, forming the basis for light-emitting diodes, Josephson junctions and bioinspired soft-hard composites. In antiferromagnet-ferromagnet (AFM-FM) heterostructures, interfacial exchange coupling can modify the coercive fields of the FM \cite{Nogues1999}, as well as provide an elusive handle to manipulate AFM order \cite{Song2018}. The former shift of the FM hysteresis curve, known as exchange bias, plays a vital role in high-density magnetic memory and read head technologies. Commonly realized in thin film AFM-FM bilayers \cite{Fukami2016,Peng2020a,Kang2021}, exchange bias has also been observed in single crystal systems that permit coexisting mesoscopic AFM and FM regions arising from chemical disorder \cite{Nayak2015,Maniv2021}. However, in these archetypal examples, the microscopic nature of the AFM-FM interface is unknown, with imperfections such as roughness, interdiffusion and stacking faults potentially decisive, preventing a rigorous understanding of the exchange coupling \cite{Nogues1999}. Particularly, the formation of domain walls near the interface is regarded as critical \cite{Stiles1999,Miltenyi2000,Scholl2004, Park2011,Noah2022}, but has not been individually visualized and controlled.

Owing to their facile hetero-integration and tunability, exchange bias with 2D magnets has been under intense pursuit \cite{Zheng2020,Gweon2021a,Chong2024,Chen2024,Zhu2020c,Albarakati2022,Ying2023,Xu2022,Phan2023}. Shifted hysteresis loops have been observed in single 2D magnets that decouple into FM and AFM slices due to perturbed interlayer interactions via intercalation \cite{Zheng2020} or surface modification \cite{Gweon2021a,Chong2024,Chen2024}, as well as in vertical heterostructures of 2D FMs with AFMs \cite{Zhu2020c,Albarakati2022,Ying2023}. However, questions linger on the definitive origin of the achieved exchange bias effects and on the reproducibility of the buried AFM-FM interfaces, which can be affected by oxidization \cite{Chong2024,Chen2024,Gweon2021a,Zhu2020c}, defect spins \cite{Xu2022} and charge transfer \cite{Phan2023}. Moreover, the dependence of interlayer exchange interactions on stacking order \cite{Song2021,Li2024} opens a new control knob, but also another source of disorder.

Here, we demonstrate a contrasting approach by utilizing the intrinsic \emph{intralayer} exchange interaction to achieve lateral exchange bias between regions of different thickness in the atomically thin, A-type AFM \CrPS{}. Since odd-layer regions possess a single uncompensated monolayer of magnetization, while even-layer regions are compensated, the connections between them create atomically sharp junctions between FM-like (odd) and AFM-like (even) regions. These pristine lateral interfaces can generate shifted odd-layer hysteresis with a single stable state at zero field due to the spring back of domain walls from boundaries with thicker even layers.

Our understanding is facilitated by quantum magnetic imaging with a single nitrogen-vacancy (NV) center spin in a scanning diamond probe (Fig. \ref{fig:1}a) \cite{Thiel2019,Song2021,Tan2024,Li2024,Tschudin2024,Palm2024}. We extend this quantitative technique to distinguish not only FM states in odd layers, but also antiphase AFM domains in even layers (e.g., $\uparrow\downarrow$ vs. $\downarrow\uparrow$) due to a weak surface magnetization. Antiphase domains in A-type AFMs have been identified previously through scanning probe techniques in bulk samples \cite{Sass2020}, but only through diffraction-limited magneto-optical techniques in few-layer samples \cite{Sun2019,Zhong2020,Qiu2023a}, precluding a detailed investigation of atomically thin, AFM domain walls. Here, we leverage high spatial resolution imaging ($\sim$60~nm) and deterministic creation of even-layer domains to elucidate the trajectory and propagation of AFM domain walls. The low magnetic anisotropy in \CrPS{} \cite{Calder2020,Peng2020} allows atomically thin even-layer regions to be switched between antiphase states at accessible magnetic fields through their weak surface magnetization, with a spread in critical fields determined by thickness, surface magnetization amplitude and interfacial exchange coupling. These features enable the simultaneous configuration of multiple AFM domains bordering the same FM-like region, which controls a multilevel lateral exchange bias.

%highlighting the potential of layered AFMs for robust, ultrathin memories.

%down to the bilayer limit 
%based on interfacial exchange coupling 

 %We achieve magnetic field control over even-layer antiphase states through their coupling to odd layers, as the reciprocal effect to exchange bias, and also through their coupling to other even layers.

\section{Tilted Magnetic Moment in Odd-Layer \CrPS{}}
Chromium thiophosphate, \CrPS, is notable among van der Waals (vdW) magnets for its desirable combination of air stability \cite{Son2021}, dispersive semiconducting bands that permit transistor operation \cite{Wu2023}, and predominant perpendicular magnetic anisotropy \cite{Calder2020,Peng2020} preferred for spintronic devices. Its crystal structure is monoclinic ($\beta = 91.9^\circ$), but close to orthorhombic, with vdW bonding along the $c$-axis (Fig. \ref{fig:1}b). Each vdW plane consists of rows of distorted, edge-sharing CrS$_6$ octahedra parallel to the $b$-axis that are separated from each other by an alternating arrangement of P atoms. The rectangular unit cell in the $ab$-plane contains four Cr atoms and is longer in the direction perpendicular to the CrS$_6$ rows ($a = 10.82$~\AA) than parallel ($b = 7.22$~\AA). Interestingly, exfoliated flakes of \CrPS{} preferentially cleave within the basal plane along the diagonal $\langle 110 \rangle$ directions of the rectangular unit cell \cite{Lee2017b}, forming characteristic parallelograms that allow identification of the $a$- and $b$-axis directions (Fig.~\ref{fig:1}c).

Neutron diffraction studies \cite{Calder2020,Peng2020} have clarified the magnetic structure of bulk \CrPS{} to consist of FM sheets that couple antiferromagnetically along the $c$-axis below a N\'eel temperature of 36 K (Fig. \ref{fig:1}b). Uniquely, the ordered Cr moment is inclined in the $ac$-plane with a projected magnitude of 2.8$\mu_B$ along the $c$-axis and $0.6\pm0.1$$\mu_B$ along the $a$-axis, corresponding to a $10\pm2^\circ$ tilt from the surface normal ($z$-axis) \cite{Calder2020,Peng2020}. For few-layer samples, recent magneto-optical Kerr \cite{Son2021} and widefield NV magnetometry experiments \cite{Huang2023} have confirmed A-type AFM order by detecting the out-of-plane moment, but whether the magnetic easy axis remains tilted is unknown.

To address this issue, we leverage the sensitivity and spatial resolution of cryogenic scanning NV magnetometry ($T = 2$ K). For NV measurements, we apply an external magnetic field $B_{ext}^{NV}$ along the NV center axis, tilted in the $xz$-plane (Fig. \ref{fig:1}a). Alternatively, for large magnetic field ramps to manipulate even-layer domains in \CrPS{}, the external field may be applied along the $z$-axis, denoted $B_{ext}^{z}$. Figure \ref{fig:1}d displays an optical image of an exfoliated \CrPS{} flake (Flake 1) displaying regions between four to six layers thick. We first focus on the 5-layer (5L) region in the bottom left corner. By scanning a single NV center spin while detecting its Zeeman-shifted magnetic resonance, we image $B_{S}$, the component of the sample's stray field  along the NV axis (\red{Methods}). For clarity, $B_S$ is always referenced to fixed laboratory coordinates (i.e., independent the sign of $B_{ext}^{NV}$).

At $B_{ext}^{NV}$ = 5~mT, the 5L region displays a negative $B_{S}$ in its interior (Fig. \ref{fig:1}e), indicating that its total magnetic moment per unit area, $\bm{\sigma} = \sum \bm{\sigma_i}$, is roughly antiparallel to the external field, where $\bm{\sigma_i}$ are the vector areal magnetizations of the individual layers. We denote this $-\sigma$ state as $\ket{-5}$, possessing layer magnetizations $\downarrow\uparrow\downarrow\uparrow\downarrow$ with the leftmost arrow denoting the bottom-most layer. To more precisely determine the moment direction, we fit the 2D image with an iterative algorithm (\red{Supplementary Sec. 4}) that assumes $\bm{\sigma}$ is uniform over the geometrically regular 5L region and oriented at polar angle $\theta_M$ from the $z$-axis and an azimuthal angle $\phi_M$ from the image horizontal ($x$-axis) (Fig. \ref{fig:1}a). %We then implement an iterative 2D fitting algorithm, which accounts for the subtle changes of the stray field along all four boundaries due to changes in $\theta_M$ and $\phi_M$.

The high data quality allows us to determine a best-fit areal magnetization $\bm{\sigma}$ for this 5L region of $-12.9 \pm 0.3$ $\mu_B$/nm$^2$ that is tilted by $\theta_M\sim16^\circ$ from the $z$-axis, with an in-plane projection oriented at $|\phi_M-\phi_a|\sim2^\circ$, where $\phi_a$ denotes the direction of the $a$-axis. This magnetization magnitude is slightly reduced from the estimated monolayer magnetization $\sigma_1 = 14.4 \pm 0.5$ $\mu_B$/nm$^2$ based on the bulk ordered moment \cite{Calder2020,Peng2020}, hinting at surface effects. The robustness of our angular determination can be gauged from the $R^2$ goodness-of-fit parameter as a function of the magnetization direction over the unit sphere (Fig. \ref{fig:1}f). Moreover, as shown in Fig. \ref{fig:1}g, the reconstructed magnetization by inverse propagation using the best-fit magnetization direction reveals an uniform magnetization over the odd-layer area. In contrast, a reconstruction that assumes a fully out-of-plane spin ($\theta_M = 0$) \cite{Thiel2019} displays spurious gradients in the magnetization, confirming the superior accuracy of the tilted spin direction for \CrPS{} (see \red{Supplementary Fig. 14}).

\section{Antiphase Domain Walls in Even-Layer \CrPS{}}
After establishing that few-layer \CrPS{} retains a tilted magnetic moment in the $ac$ plane, we now discover that antiphase domains in even-layer regions can be distinguished by high-sensitivity NV magnetometry. Figure \ref{fig:2}a displays the $B_S$ image for Flake 1 over a larger field-of-view. Here, the bottom 5L region adopts the $\ket{-5}$ state, while the top 5L stripe adopts $\ket{+5}$. A weaker contrast is observed in the 4L and 6L regions, with the area closer to the bottom (top) 5L displaying a positive (negative) stray field. Normally, we may expect an even-layer region to possess layer magnetizations aligned with those in the odd-layer region connected to it (i.e., the 4L and 6L regions adjacent to $\ket{-5}$ take the states \mfour{} and \msix{}, denoted as $\ket{-4}$ and $\ket{-6}$, respectively). Indeed, our contrast here stems from an antiphase domain wall in the 6L (higher resolution measurement shown in Fig. \ref{fig:2}b), and the antiphase even-layer domains stabilize their adjoining odd-layer regions into opposite states.

Before exploring this apparent exchange bias, we examine the origin of the magnetic contrast in even-layer \CrPS{}, which differs from the zero magnetization for even layers reported by prior magnetometry experiments on A-type AFMs \cite{Thiel2019,Huang2023,Tschudin2024,Zur2023,Gao2024,Zhang2024}. By imaging the edge between the 4L and 6L regions (black dashed line in Fig. \ref{fig:2}d), we find no noticeable transition in stray field, indicating that our signal arises from a surface, rather than volume magnetization (see also Fig. \ref{fig:2}g).

In Fig. \ref{fig:2}f, we inspect another flake (Flake 2) containing two 3L regions bridged by 2L and 4L regions. At $B^{NV}_{ext}$ = $-10$~mT, the 3L regions exist in opposite states due to exchange bias, and a ridge in the bilayer is faintly perceptible. A finer magnetic resolution image, shown in Fig. \ref{fig:2}g, conclusively resolves the antiphase domain wall in the bilayer. Intriguingly, the $\ket{-2}$ (\mtwo) region displays a negative stray field, while the $\ket{+2}$ (\ptwo) and $\ket{+4}$ (\pfour) regions display positive stray fields. The sign of these fields is opposite to Flake 1, where $\ket{+}$ and $\ket{-}$ even-layer states display negative and positive stray field, respectively.

Our observations indicate that the stray field in even-layer \CrPS{} should arise due to a difference in the magnetizations of the top and bottom surface monolayers. Differences in the distances between different layers to the NV center fly height $z_{NV}$ can contribute a nonzero signal even for compensated even layers; however, we find that this signal is negligible for $z_{NV} = 60$~nm (\red{Supplementary Sec. 6}).  Surface layers of vdW magnets can possess different properties from bulk layers due to missing exchange links \cite{Tschudin2024,Guo2024}, structural relaxation \cite{Guo2024}, surface reactivity \cite{Chong2024,Zhang2024}, electric fields and charge transfer \cite{Jiang2018}. In human-made samples, such changes need not be identical for the top and bottom surfaces. Thus, a probe sensitive to both surfaces and possessing sufficient sensitivity should distinguish antiphase domains. Evidently, in Flake 1, the magnetization of the top layer should be larger than that of the bottom layer ($\sigma_{top}>\sigma_{bot}$) (Fig. \ref{fig:2}e), while the opposite ($\sigma_{top} < \sigma_{bot}$) is true for Flake~2 (Fig. \ref{fig:2}h). Data on five flakes support a stochastic even-layer signal with $|\sigma_{even}| = |\sum \sigma_i| \approx |\sigma_{top}+\sigma_{bot}|$ between 0.4 to 2.2 $\mu_B$/nm$^2$ that arises predominantly from a difference in the magnitude, rather than direction, of the surface layer magnetizations (\red{Supplementary Figs. 1 and 3}). This surface magnetization ($\sim$1 $\mu_B$/nm$^2$) is comparable in magnitude to that observed in the bulk AFM Cr$_2$O$_3$ at room temperature \cite{Hedrich2021,Makushko2022,Wornle2021}. However, due to atomic thinness here, it is contributed by both surfaces and crucially enables, as we will show, control over the entire AFM state, including inner layers.

This empirical contrast opens a stunning window into the dynamics of atomically thin AFM domain walls. The trajectory of the antiphase domain wall is governed by the minimization of the full micromagnetic free energy, although energy barriers may constrain the trajectory to a local minimum. This leads to a preference for shorter domain wall lengths and for domain walls to locate in thinner layers and intersect free edges perpendicularly (i.e., $(\hat{n}\cdot\nabla)\bm{M} = 0$, where $\hat{n}$ is the normal to the boundary and $\bm{M}$ is the volume magnetization \cite{Hubert1998}).

For both Flake 1 and 2, the shortest domain wall trajectory, favored at zero field, is a straight line perpendicular to the free edge (e.g., left panel in Fig. \ref{fig:2}c). Moreover, the trajectories intersect odd-layer corners, which can represent local minima in energy due to geometric factors and dipolar coupling. For example, when the wide 3L in Flake 2 is $\ket{-3}$ (Fig. \ref{fig:2}f), the domain wall traces through its corner, since a shift left increases the domain wall length (adds top boundary of $\ket{-3}$), while a shift right moves it into the thicker 4L region. At larger $B^z_{ext}$, the domain wall curves outwards with one end fixed at the odd-layer corner to increase the area of the domain whose uncompensated magnetization is aligned with $B^z_{ext}$. This tendency of even-layer domain walls to intersect and pivot about odd-layer corners is observed in both our continuum micromagnetic (Fig. \ref{fig:2}c,i) and atomistic spin dynamics simulations (\red{Supplementary Secs. 10 and 11}).

\section{Exchange Bias and Asymmetric Magnetization Reversal Mechanisms}
We turn to investigating the exchange bias on odd-layer regions. For \CrPS{}, its weak magnetocrystalline anisotropy per ion $K$ is dominated by the AFM interlayer exchange coupling $J_c$ ($K/J_c \approx 0.025$) \cite{Calder2020,Peng2020,Shi2024}. Hence, a spin-flop transition $B_{flop}$ ($\propto \sqrt{J_c K} \approx 0.8$ T) takes precedence over spin-flip transitions ($\propto J_c$) \cite{Wang2019b,Shi2024}. For atomically thin samples, a coercive (anisotropy) field $B_c$ where the magnetization of all layers simultaneously reverses to attain the minimum energy antiphase state (e.g. \mthree{} to \pthree) can potentially be observed below the spin-flop transition at $B_c \propto N K / \sigma$, where $N$ is the number layers and $\sigma$ is the net magnetization for all layers. Here, the anisotropy barrier against flipping all layers must be overcome by the Zeeman energy reduction for the largely compensated magnetization (e.g., for odd-layers, $\sigma \approx \sigma_1$), implying that $B_c$ will not be observed for the bulk, above a critical $N$.

We first verify that the magnetic hysteresis for an isolated 3L region is symmetric, displaying coercive fields $B^{NV}_c = \pm19$~mT for fields along the NV axis and $B^{z}_c = \pm95$~mT for fields along the $z$-axis \red{(Supplementary Fig. 4}). A reduction in coercive field is expected in the Stoner-Wohlfarth model for fields making larger angles to the anisotropy axis (i.e., $|B^{NV}_{c}| < |B^{z}_{c}|$ for \CrPS{}) \cite{Blundell2001}. In contrast, the hysteresis curves for the two 3L regions in Flake 2 (denoted narrow and wide) that are surrounded by even-layer regions display a prominent asymmetry (Fig. \ref{fig:3}a). With all even layers initialized as $\ket{+}$ states, the hysteresis loops for the narrow and wide 3L are both offset to negative $B^{NV}_{ext}$; however, since their loop areas differ from each other and from the isolated 3L (shown as the red background rectangle), the exchange bias cannot be interpreted as a simple shift. This reveals different reversal mechanisms at the two coercive fields \cite{Stiles1999,Beckmann2003}. We define $B_{c2}$ as the upper coercive field required to switch the odd-layer magnetizations antiparallel to the even-layer regions  (i.e., $\ket{+3}\rightarrow\ket{-3}$ here) and $B_{c1}$ as the lower coercive field needed to switch them back to alignment (i.e., $\ket{-3}\rightarrow\ket{+3}$ here).

To understand how the magnetization reverses, we perform detailed imaging of both 3L regions (Fig. \ref{fig:3}b-f). As $B^{NV}_{ext}$ decreases from the positive side to $-41$~mT (between the upper coercive fields $B_{c2}$ for the wide and narrow 3L), the wide 3L flips to $\ket{-3}$, while the narrow 3L remains exchange-biased as $\ket{+3}$ (Fig. \ref{fig:3}b). Decreasing $B^{NV}_{ext}$ further to $-70$~mT flips the narrow 3L to $\ket{-3}$, and now domain walls encircle both 3L regions (Fig. \ref{fig:3}c). The reversal at $B_{c2}$ occurs instantaneously over the complete area of each odd-layer region. Hence, it proceeds by a coherent rotation of the odd-layer moments, which winds a lateral exchange spring at the odd-even layer interface, in analogy to planar domain walls in conventional AFM-FM bilayers \cite{Scholl2004, Park2011}. Reflective magnetic circular dichroism (RMCD) measurements over a wider field range confirm that the antiphase transition ($B_c$) for a 3L region is followed by a spin-flop transition at $|B^z_{ext}|\approx1$~T \red{(Supplementary Fig. 5)}.

In contrast, when ramping $B^{NV}_{ext}$ towards positive fields, the narrow 3L switches back to $\ket{+3}$ at $-2$ mT, \emph{prior to} encountering positive field, while the edge between the wide 3L and the $\ket{+4}$ region becomes magnetically ragged (Fig. \ref{fig:3}d), foreshadowing the depinning of the domain wall. We can also image the narrow 3L just before it switches to confirm that its $B_{c1}$ reversal is similarly caused by domain wall motion (Fig. \ref{fig:3}f). Moreover, the precise trajectory of the domain wall during reversal is determined by the configuration of the surrounding even-layer domains. In Fig. \ref{fig:3}g, we initialize the even layers into a domain state, where the right region is $\ket{-4}$, while the bottom and left regions are $\ket{+4}$ and $\ket{+2}$, respectively. In this case, the dominant exchange bias comes from the longer and thicker interface to $\ket{-4}$, and the $B_{c1}$ reversal occurs by horizontal translations of the domain wall away from this long interface, in contrast to vertical translations in the uniform state (Fig. \ref{fig:3}f).

These phenomena can be understood from micromagnetic simulations, which reveal that the domain wall formed at $B_{c2}$ between an odd layer and a thicker even layer (e.g., right edge in Fig. \ref{fig:3}f) presses against their interface, while that between an odd layer and a thinner even layer (e.g., left edge in Fig. \ref{fig:3}f) escapes to the even layer. As the field amplitude is reduced towards $B_{c1}$, the domain wall at a thicker even-layer edge can spring back across the odd-layer with negligible energy barrier, truncating roughly half of the hysteresis curve ($B_{c1} \approx 0$). Specifically, for enclosed geometries (e.g. Fig. \ref{fig:3}f), the surface tension of domain walls causes the domain wall bubble to collapse prior to crossing zero field \red{(see Supplementary Sec. 7 for a theoretical model)}. In this case, $B_{c1}$ occurs on the same side as $B_{c2}$ (e.g., at $-2$~mT for reversal to $\ket{+3}$ in Fig. \ref{fig:3}f), and only the odd-layer state aligned with the even layers is stable at zero field. Alternatively, for geometries where the reversal process largely maintains domain wall length, reversal requires slight field pressure to overcome pinning and $B_{c1}$ occurs just after crossing zero field (e.g., at $-1$~mT for reversal to $\ket{-3}$ in Fig. \ref{fig:3}g). However, this spring back effect generally does not occur for interfaces between an odd layer and thinner even layer \cite{Son2021}, as the domain wall that forms on the even-layer side is ideally field-insensitive and is energetically impeded from re-entering a thicker region.

To verify the domain wall locations, we perform stray field linecuts across the narrow 3L both when it is antialigned (e.g., Fig. \ref{fig:3}c) and aligned (e.g., Fig. \ref{fig:3}e) with the surrounding even layers. We constrain the magnetization magnitude ($\sigma = 11.0$ $\mu_B$/nm$^2$) and NV center tip height ($z$ = 60~nm) from the linecut without domain walls (Fig. \ref{fig:3}i). We then use these parameters to fit the linecut with domain walls at both edges (Fig. \ref{fig:3}h), but assume a stepwise reduction of the magnetization to zero outside the odd layer. The deviation (green trace in Fig. \ref{fig:3}h) between the data and this simplified fit then visualizes the effect of the unaccounted domain wall texture. Indeed, we find minimal discrepancy between the data and model at the left edge (between $\ket{+2}$ and $\ket{-3}$), confirming that the domain wall lies inside the thinner bilayer and thus is virtually compensated. On the other hand, the data at the right edge (between $\ket{-3}$ and $\ket{+4}$) displays a substantially weaker and wider signal than modeled, indicating an extended rotation of the magnetization on the 3L side upon approaching the interface \cite{Scholl2004, Park2011}.

The reversal at the upper coercive field $B_{c2}$ balances the energy cost to nucleate domain walls along the perimeter of the odd layer with the Zeeman energy gain from aligning its magnetization with the field. Thus, $B_{c2}$ should be shifted by an additional exchange field ${B_{E} \propto  \varepsilon_{DW} C t_{e}/ (2\sigma_{1} S)}$ (\red{Methods}), where $S$ is the area of the odd-layer region. The numerator evaluates the total energy of the domain walls, given by the product between the domain wall energy density $\varepsilon_{DW}$ and the total cross-sectional area of the walls, with $C$ as the perimeter and $t_e$ as the effective thickness of the even layers surrounding the odd layer. The lateral exchange bias, proportional to the perimeter-to-area ratio $C/S$, thus scales roughly as the inverse of the width of the odd-layer region (dimension perpendicular to the dominant lateral interface). This is analogous to the inverse scaling of the exchange bias with the thickness of the FM in vertical AFM-FM bilayers \cite{Nogues1999}; however, here interfacial coupling can be much stronger as it is supplied by the intralayer exchange within a single material. Indeed, the shift in $B_{c2}$ relative to the isolated 3L is larger for the narrow 3L ($B_{E}^{NV} = -52$~mT) than for the wide 3L ($B_{E}^{NV} \approx 0$~mT) (Fig. \ref{fig:3}a). Notably, exchange fields $B_E^{NV}$ along the NV axis are smaller than $B_E^{z}$ along the $z$-axis (\red{Supplementary Fig. 6}), consistent with generally decreasing exchange bias observed as the field rotates from the anisotropy axis \cite{Ambrose1997,Xi1999}. Moreover, due to AFM interlayer coupling, $B_E$ increases with thickness $t_e$ ($\propto N$, the number of layers). Indeed, for the 5L corner region in Flake 1 (Fig. \ref{fig:1}e), both $B_{c2} = -151$~mT and $B_{c1} = -3$~mT increase in magnitude over the 3L, reflecting stronger interfacial exchange coupling (Fig. \ref{fig:3}a). 

Our lateral exchange bias is related to effects observed recently in extrinsically patterned thin films of amorphous ferrimagnetic alloys \cite{Liu2023}. As advantages, our approach based on layer parity in intrinsic 2D A-type AFMs offers inherent magnetic compensation in AFM (even-layer) regions without fine-tuning, pinning of domain walls at \emph{atomically sharp} thickness gradations (spring back at $B_{c1}\approx 0$), and scaling of the exchange bias with thickness.

\section{Control over Antiphase Even-Layer Domains}
The antiphase states of even-layer regions in \CrPS{} can be manipulated by magnetic field and interfacial effects similar to odd layers. Since their uncompensated surface magnetization $\sigma_{even}\ll \sigma_1$, even layers should display antiphase switching fields $B_c \propto N K / \sigma_{even}$ that are significantly higher than comparably thick odd layers. Direct imaging of antiphase switching above $B_{c2}$ is thus challenging for thicker even layers ($N > 2$) due to the limited field range of NV magnetometry.

We identify a large 2L region in Flake 3 (Fig. \ref{fig:4}a inset) that borders a 4L region. The $\ket{+2}$ and $\ket{+4}$ states display positive surface magnetizations (Fig. \ref{fig:4}b). After ramping $B^{NV}_{ext}$ to -124 mT, the bilayer flips to $\ket{-2}$, aligning its weak surface magnetization with negative field (Fig. \ref{fig:4}c). The bilayer hysteresis with fields smaller than the 4L coercive field displays an asymmetry analogous to odd layers (Fig. \ref{fig:4}a). Due to reduced thickness and small boundary-to-area ratio here, interfacial coupling to $\ket{+4}$ should contribute only a weak exchange field $B_E$ at $B_{c2}$; however, it strongly reduces $B_{c1}$ by allowing reversal by domain wall motion. As visualized in Fig. \ref{fig:4}d, the reversal from $\ket{-2}$ to $\ket{+2}$ occurs by domain wall translation from the 4L interface at $B^{NV}_{c1} = 17$~mT ($\ll |B^{NV}_{c2}| = 102$ ~mT).

Remarkably, surface magnetization can also drive the antiphase switching of thicker even layers. Figure \ref{fig:4}e displays a fourth \CrPS{} flake (Flake 4), containing consecutive 4L and 6L regions that display negative surface magnetizations in their $\ket{+}$ states, opposite to Flake 3. In this case, we expect positive $B^z_{ext}$ to reverse $\ket{+}$ to $\ket{-}$. Indeed, ramping $B^z_{ext}$ to 0.35 T and back near zero, the 4L region and the 6L region connected to the 4L switch together to $\ket{-}$ (Fig. \ref{fig:4}f). The 6L coercive field is reduced by interfacial coupling to the 4L, since an isolated 6L should have higher coercive field, consistent with the absence of switching for the disconnected portion of the 6L (left side). A larger ramp to 1.0~T ($\hat{z}$) flips the remaining $\ket{+6}$ region to $\ket{-6}$ (Fig. \ref{fig:4}g), while a negative field ramp to $-0.3$ T toggles the connected 4L and 6L regions back to $\ket{+}$ (Fig. \ref{fig:4}h).

Furthermore, we can use magnetic field pulses to control the location of the even-layer antiphase domain wall. In Fig. \ref{fig:4}i (center panel), we nucleate a domain wall in the 6L region of Flake 1 by starting from uniform $\ket{+}$ states (inset) and pulsing $B^z_{ext}$ to 0.6~T to reverse the connected 4L and 6L regions to $\ket{-}$ by their surface magnetization. The upper half of the 6L does not flip to $\ket{-}$ at the same time due to interfacial exchange bias from the top odd and bulk regions, which adopt the opposing $\ket{+}$ state at positive field.

In A-type AFMs, a magnetic field drives the domain wall in opposite directions in the two layer sublattices \cite{Hadjoudja2024}. Hence, symmetry breaking is required for unidirectional propagation. For our weakly uncompensated even layers, $B^z_{ext}$ at the apex of the ramp drives the wall to increase the area of the domain with $\sigma$ aligned with the field, potentially increasing the length of the wall (Fig. \ref{fig:4}i). Upon returning to near zero field, the domain wall remains in place or only partially retreats stochastically due to pinning. These metastable wall trajectories are reproduced by snapshots of the field-induced domain wall evolution in micromagnetic simulations that assume an uncompensated even-layer magnetization (Fig. \ref{fig:2}c and \red{Supplementary Fig. 19}).

\section{Reconfigurable, Multilevel Exchange Bias}
We finally demonstrate discrete-level control over the exchange bias amplitude by toggling the even-layer antiphase domains surrounding an odd layer. For Flake 2 (Fig. \ref{fig:2}f), we can achieve any arbitrary combined state $\ket{2\!\cdot\!4_B\!\cdot\!4_R}$ for the three even-layer regions, labeled $\ket{2}$, $\ket{4_B}$, $\ket{4_R}$, that border the narrow 3L on its left, bottom, and right sides (Fig. \ref{fig:5}a). This is possible when different regions possess distinct coercive fields $B_c$. Ramping below a negative critical field $B_{ext}^z \approx -0.3$~T, we switch the bilayer from $\ket{+2}$ to $\ket{-2}$ through its surface magnetization (see Fig. \ref{fig:2}g,h). In contrast, the 4L regions, $\ket{4_R}$ and $\ket{4_B}$, switch from $\ket{+4}$ to $\ket{-4}$ above different \emph{positive} critical fields ($B_{ext}^z = 0.6$~T and 0.8 T, respectively). This behavior is ostensibly opposite to a surface magnetization-driven switching. Indeed, we reveal below that the 4L regions possess a particularly weak surface magnetization and an atypical edge magnetization.

In Fig. \ref{fig:5}b, we start in $\ket{-\!-\!-}$ and subsequently generate in order $\ket{-\!-\!+}$, $\ket{+\!-\!+}$, and $\ket{-\!+\!+}$ by three judiciously sized field pulses. The time-reversed copies of these four states can also be realized by first creating $\ket{+\!+\!+}$, for example through a small positive field ramp from the final $\ket{-\!+\!+}$ state. In Fig. \ref{fig:5}c, we image the resulting antiphase domains in the $\ket{2}$, $\ket{4_R}$ and $\ket{4_B}$ regions, recalling that in Flake 2, $\ket{+}$ states have positive stray fields $B_S$. The signal interior to the 4L is inhomogeneous and overshadowed by a strong edge signal, where the latter also correlates with the antiphase state. This edge signal (apparent also in the linecuts of Fig. \ref{fig:3}h,i) is not only stronger, but also the opposite sign to that expected for a simple termination of the interior magnetization (c.f. right edges of odd layers with similar geometry in Fig. \ref{fig:3}). It likely results from an uncompensated edge magnetization that is opposite to the interior, such as due to a sloped cleave, and could instead determine switching (\red{see Supplementary Sec. 8}).%, . %or edge anisotropy.

Figure \ref{fig:5}d presents hysteresis curves for the narrow 3L after configuring $\ket{2\!\cdot\!4_B\!\cdot\!4_R}$ into different states. The upper coercive field $B^{NV}_{c2}$ for $\ket{-\!-\!-}$ is naturally opposite in sign and similar in magnitude (71 mT) to that for $\ket{+\!+\!+}$ (Fig. \ref{fig:3}a). In comparison to $\ket{+\!+\!+}$, as we progressively introduce additional interfaces to $\ket{-}$ regions, $|B^{NV}_{c2}|$ is reduced to 56 mT after flipping only $\ket{2}$ ($\ket{-\!+\!+}$), to 34~mT after flipping only $\ket{4_B}$ ($\ket{+\!-\!+}$) and to 27 mT after flipping both $\ket{2}$ and $\ket{4_B}$ ($\ket{-\!-\!+}$). In all, we can encode $2^3$ different levels of exchange bias through the $\ket{2\!\cdot\!4_B\!\cdot\!4_R}$ register.

\section{Discussion}
Using high-sensitivity, high-spatial-resolution quantum magnetometry, we directly visualized how the training of AFM domains with magnetic field governs the lateral exchange bias in atomically thin \CrPS{}. Our imaging contrast arises from a weak surface magnetization, which additionally enables control over the full even-layer state, though its origin requires further study. These advancements in imaging and control will stimulate research on manipulating AFM domain walls by electric fields, spin orbit torques and strain. Particularly, bilayer domain walls in \CrPS{} in proximity to a heavy metal may achieve enhanced velocities under current-induced chiral spin torques as observed in synthetic AFMs \cite{Yang2015a}, opening an alternative, single-crystal platform for AFM racetrack devices. Our tunable exchange bias combined with in-plane electrical transport \cite{Wu2023} further suggests a new architecture for a lateral spin valve memory, where two odd-layer regions with different exchange bias amplitudes (one ``free'' and one ``pinned'') are linked by an even layer (\red{Supplementary Fig. 9}). Finally, the principles for visualization and interfacial control here may be extensible to diverse A-type 2D AFMs, fostering the exploration of AFM skyrmions \cite{Tan2024} and topological chiral edge states at domain walls \cite{Qiu2023a}.
%The insights presented are enabled by our visualization of antiphase even-layer states in \CrPS{} by high-sensitivity, high-spatial-resolution NV magnetometry. 

\paragraph*{Note added.\textemdash}
Recently, lateral exchange effects were shown to govern the layer flipping sequence of connected even layers in another 2D A-type AFM, CrSBr \cite{Sun2024, PelletMary2025}.

%\bibliography{C:/Users/zhouqt/GoogleDrive/BibTex/0_CrPS4}

%\clearpage{}
%\newpage
\setcounter{section}{0}
%\begingroup
%\raggedbottom

\section{Acknowledgments}
The authors thank Y. Ran, P. Aynajian for valuable discussions and Q. Ma, K. S. Burch for use of some laboratory facilities. B.B.Z. and Y.-X.W. acknowledge support from the Department of Energy Early Career Program under award number DE-SC0024177 for NV center measurements. B.B.Z., Y.-X.W. and T. K. M. Graham acknowledge support from the National Science Foundation (NSF) award DMR-2047214 for development of the scanning NV microscope. B.B.Z. and X.-Y.Z. were supported by NSF ECCS-2041779 for sample fabrication and magnetization analysis. Z.L., M.A.I, R.N.G., G.P.T. and T.A. acknowledge a startup fund from Binghamton University. Funding for the ADL Small Grants Program is made possible by support to S3IP from New York Empire State Development Division of Science, Technology, and Research. E.J.G.S. acknowledges computational resources through CIRRUS Tier-2 HPC Service (ec131 Cirrus Project) at EPCC (http://www.cirrus.ac.uk), which is funded by the University of Edinburgh and EPSRC (EP/P020267/1); and ARCHER2 UK National Supercomputing Service via the UKCP consortium (Project e89) funded by EPSRC grant ref EP/X035891/1. E.J.G.S. acknowledges the EPSRC Open Fellowship (EP/T021578/1) and the Donostia International Physics Center for funding support. E.J.G.S. and R.R.-E. acknowledge support from the Royal Society through the International Newton Fellowship (NIF/R1/241532). K.W. and T.T. acknowledge support from the JSPS KAKENHI (Grant Numbers 21H05233 and 23H02052) and World Premier International Research Center Initiative (WPI), MEXT, Japan.

\section{Author contributions}
Y.-X.W., Z.L. and B.B.Z. conceived the experiments. M.A.I., G.P.T. synthesized the bulk crystals; M.A.I., R.N.G., G.P.T. and T.A. performed bulk crystal characterizations and RMCD measurements. C.B. performed x-ray diffraction measurements. Z.L. supervised the research at Binghamton University. Y.-X.W. and T.K.M.G. developed the scanning NV magnetometry instrumentation and protocols. Y.-X.W. fabricated the samples and performed the scanning NV experiments, with the assistance of T.K.M.G. and X.-Y.Z. Y.-X.W. analyzed the data and performed micromagnetic simulations, with the assistance of X.-Y.Z. and B.B.Z. Atomistic spin dynamics simulations were performed by R.R.-E., and the Stoner-Wohlfarth model was developed by M.B., both under the guidance of E.J.G.S. K.W. and T.T. synthesized the hBN crystals. B.B.Z., Y.-X.W., R.R.-E., M.B. and E.J.G.S. wrote the manuscript, with input from all authors.

\section{Competing interests}
The authors declare no competing interests.

%\clearpage{}
\section{Figure Captions}

%FIGURE 1
\begin{figure*}[hbt]
\includegraphics[scale=1]{Fig1.pdf}
\caption{\label{fig:1}Layered AFM in atomically thin \CrPS{}. (a) A single NV center inside a diamond probe is scanned over few-layer \CrPS{}. The parity ($\ket{+}$ or $\ket{-}$) of an odd layer is determined by the direction of its uncompensated layer of magnetization. Even-layer and odd-layer states with the same parity have aligned layer magnetizations. The direction of the net areal magnetization $\bm{\sigma}$ is specified by a polar angle $\theta_M$ and an in-plane angle $\phi_M$ ($x$-axis). For magnetic imaging, the external field $B_{ext}^{NV}$ is applied at 54.7$^{\circ}$ from normal in the $xz$-plane, parallel to the NV axis. (b) The magnetic moments (red and blue arrows) in bulk \CrPS{} alternate between adjacent layers (A-type AFM) and is tilted in the $ac$-plane. (c) Crystallographic structure in the $ab$-plane, showing two inequivalent Cr atoms within the rectangular unit cell. A row of edge-sharing CrS$_6$ octahedra along the $b$-axis is highlighted in blue. Exfoliated flakes typically form parallelograms bounded by the [110] directions. (d) Optical image of Flake 1 with layer thicknesses labeled. (e) Image of the sample stray field $B_S$ along the NV center axis for the 5-layer corner region in d). The cartoon (inset) shows the layer magnetizations for the $\ket{-4}$, $\ket{-5}$, and $\ket{-6}$ states. (f) Goodness-of-fit parameter $R^2$ for fitting e) using a uniform $\bm{\sigma}$ over the 5L area, oriented along arbitrary directions over the unit sphere. (g) Reconstructed magnetization amplitude $\sigma$ using the best-fit direction, determined to be tilted near the $ac$-plane.}
\end{figure*}

%FIGURE 2
\begin{figure*}[hbt]
\includegraphics[scale=1]{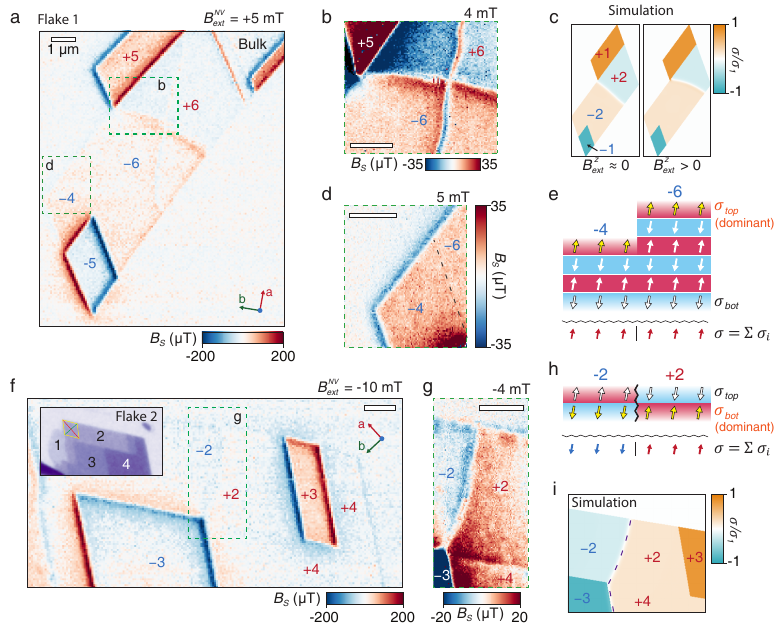}
\caption{\label{fig:2}Antiphase domains in even-layer \CrPS{}. (a) Stray field $B_S$ image of Flake 1. The top and bottom 5L regions adopt opposite magnetization states $\ket{+5}$ and $\ket{-5}$, respectively. The weak $B_S$ in the intervening even-layer region reveals an antiphase domain wall between $\ket{+6} =$ \psix{} above and $\ket{-6} =$ \msix{} below. (b) Higher resolution $B_S$ measurement of the 6L antiphase domain wall. The vertical ridge in $B_S$ corresponds to a physical defect (fold) in the sample. (c) Micromagnetic simulation of even-layer domain wall trajectories for a representative geometry. For small positive $B_{ext}^z$, the domain wall takes a minimum length state that touches the odd-layer corner (left). For larger positive $B_{ext}^z$, the domain wall curves to increase the area of the $\ket{-}$ even-layer domain, while avoiding the odd-layer interface (right). (d) $B_S$ image of the transition between 4L and 6L, demonstrating that the even-layer signal is not proportional to thickness. (e) The even-layer $B_S$ plausibly arises from a deviation from perfect compensation between the areal magnetizations $\sigma_{top}$ and $\sigma_{bot}$ of the top and bottom surface layers. The net magnetization $\sigma$, summed over all layers, is shown beneath the wavy line. For Flake 1, the data imply $\sigma_{top} > \sigma_{bot}$. (f) $B_S$ image of \CrPS{} Flake 2 (optical image in inset). (g) Higher resolution $B_S$ image of the domain wall in the bilayer, with $\ket{-2}$ = \mtwo{} on the left and $\ket{+2}$ = \ptwo{} on the right. (h) For Flake 2, the sign of $B_S$ for its even-layer states suggests $\sigma_{bot} > \sigma_{top}$, opposite to Flake 1. (i) Simulation for the bilayer domain wall in Flake 2. The simulations in c) and i) assume uncompensated even-layer magnetizations with the same sign as experiment. All scale bars denote 1 $\mu$m.}
\end{figure*}

%FIGURE 3
\begin{figure*}[hbt]
\includegraphics[scale=1]{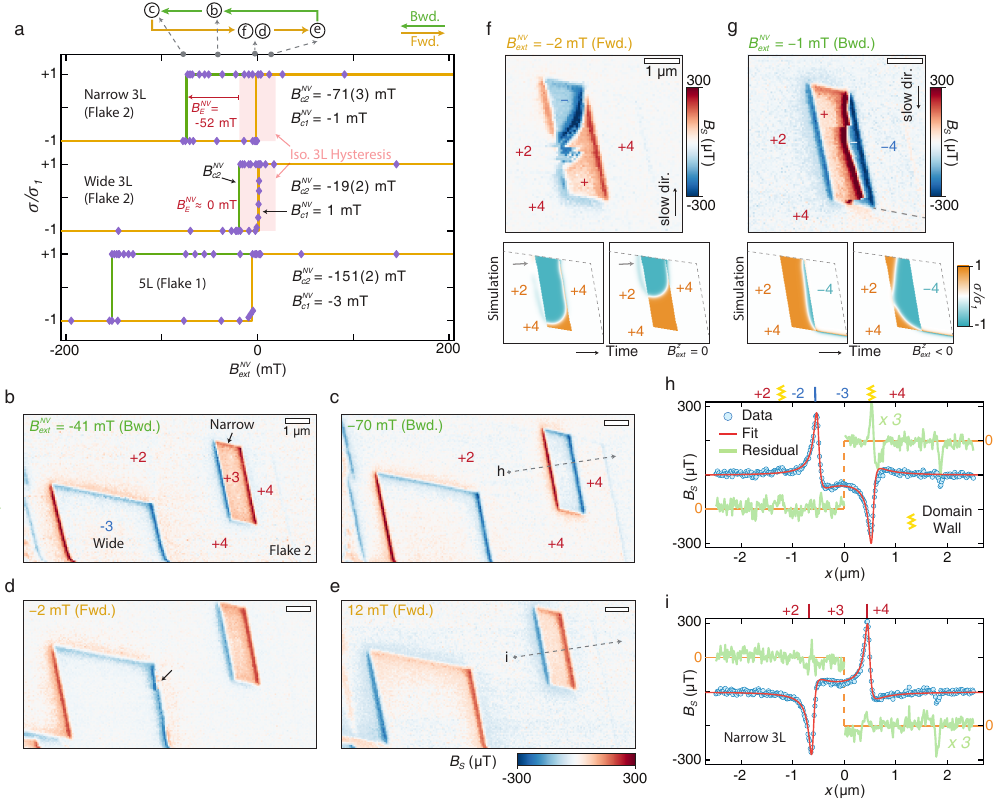}
\caption{\label{fig:3}Lateral exchange bias at the even-odd interface. (a) Hysteresis curves for the narrow and wide 3L regions in Flake 2 and the corner 5L in Flake 1 with the surrounding even-layer regions in uniform $\ket{+}$ states. For comparison, the symmetric hysteresis loop for an isolated 3L region is shown as the red background rectangle. (b-e) Stray field $B_S$ images of the narrow and wide 3L at various $B^{NV}_{ext}$ on the hysteresis curve (labeled in a). Domain walls that encircle the wide 3L in the $\ket{-3}$ state begin to depin on the forward sweep prior to $B^{NV}_{ext} = 0$ (see arrow in d). Bwd. - backward sweep; fwd. - forward sweep. (f) Image of the magnetization reversal at $B_{c1}$ for the narrow 3L when all even layers are $\ket{+}$. Barkhausen jumps are thermally activated by scanning (slow scan direction indicated by arrow).  Simulations (bottom) reveal that the domain wall between 2L and 3L occurs on the bilayer side (arrow), while that between 3L and 4L is pinned at the interface. (g) Data (top) and simulations (bottom) of the $B_{c1}$ reversal when nearby even layers are in a domain state, as labeled. (h) $B_S$ linecut across the narrow 3L when domain walls exist along both edges (e.g. panel c). (i) $B_S$ linecut when no domain walls exist (e.g., panel e). The green trace (residual) in h) or i) presents the deviation of the data from a fit assuming no domain walls. For clarity, residuals are multiplied by three and zero residual is offset differently for $x < 0$ and $x > 0$.}
\end{figure*}

%FIGURE 4
\begin{figure*}[hbt]
\includegraphics[scale=1]{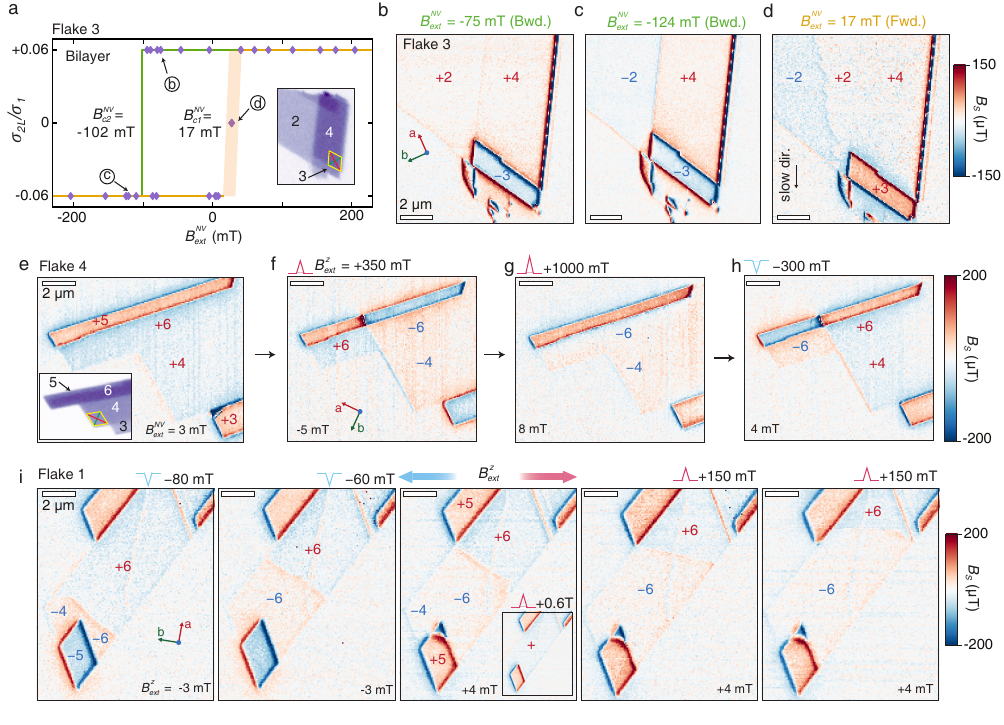}
\caption{\label{fig:4}Controlled nucleation and translation of AFM domain walls. (a) Hysteresis loop for a bilayer region in a third \CrPS{} flake (Flake 3, optical image in inset). The bilayer hysteresis is driven by its uncompensated surface magnetization and is asymmetric due to interfacial coupling to the 4L region on its right. (b,c,d) $B_S$ images of Flake 3 at various points on the bilayer hysteresis loop, as labeled in a). Large negative $B^{NV}_{ext}$ flips the bilayer from $\ket{+2}$ (with $+\sigma$) to $\ket{-2}$ (with $-\sigma$), while a small positive field reverses $\ket{-2}$ to $\ket{+2}$ by domain wall translation. (e) $B_S$ image of a fourth \CrPS{} flake (Flake 4) with all regions in $\ket{+}$ states. A narrow 6L stripe is partially connected to a wider 4L region (inset). (f) Image after ramping $B^z_{ext}$ to 0.35~T and back. The connected 4L and 6L areas switch from $\ket{+}$ to $\ket{-}$. (g) Image after ramping $B^z_{ext}$ to 1~T and back. The remaining $\ket{+6}$ domain on the left switches to $\ket{-6}$. (h) Image after ramping $B^z_{ext}$ to $-0.3$~T and back. The connected 4L and 6L areas now reverse from $\ket{-}$ to $\ket{+}$. (i) Field-driven motion of even-layer domain walls in Flake 1. The initial domain state (center) is formed from a uniform $\ket{+}$ state (inset) by ramping $B^{z}_{ext}$ to +0.6 T. The domain wall can then be driven left or right by negative or positive field pulses, respectively.}
\end{figure*}

%FIGURE 5
\begin{figure*}[hbt]
\includegraphics[scale=1]{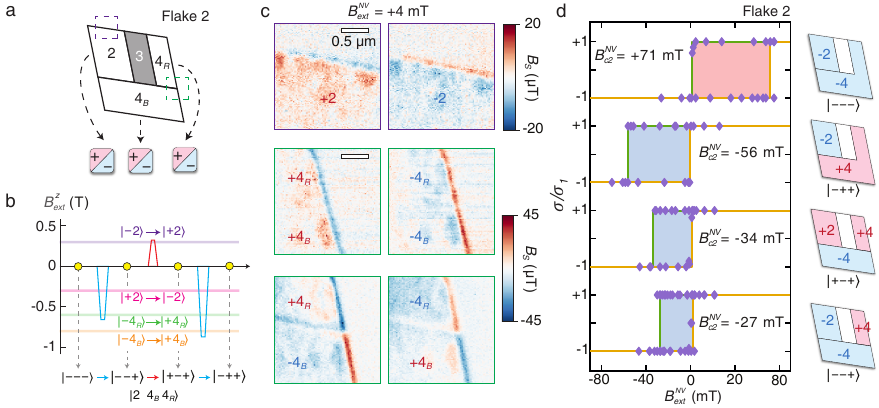}
\caption{\label{fig:5}Multilevel exchange bias through digital control of AFM domains. (a) Outline of the area surrounding the narrow 3L in Flake 2. Three independent even-layer domains can be defined: bilayer ($\ket{2}$), bottom 4L ($\ket{4_B}$), and right 4L ($\ket{4_R}$), forming a composite three-bit state $\ket{2\!\cdot\!4_B\!\cdot\!4_R}$. (b) Pulse sequence for generating different configurations of the three even-layer domains, starting from $\ket{2\!\cdot\!4_B\!\cdot\!4_R} = \ket{-\!-\!-}$. The horizontal colored lines denote the critical fields for the transition labeled in the same color. For example, a pulse to $-0.8$~T exceeds all three negative critical fields and generates the state $\ket{-\!+\!+}$, regardless of the initial state. (c) $B_S$ images of the even-layer domains in various states. Imaged regions depicted by boxes in a). The sign and strength of the stray field on the 4L edge is incompatible with a termination of the interior magnetization and suggests different magnetic uncompensation at the edge. (d) Hysteresis loops displaying a tunable $B^{NV}_{c2}$ for the narrow 3L when the even-layer domains are initialized as $\ket{-\!-\!-}$, $\ket{-\!+\!+}$, $\ket{+\!-\!+}$, and $\ket{-\!-\!+}$ from top to bottom.}
\end{figure*}

\clearpage{}
\section{References}
\bibliography{0_CrPS4}

\section{Methods}
\subsection{Measurement details}
Bulk \CrPS{} crystals were synthesized using chemical vapor transport with iodine as the transporting agent \cite{Budko2021}. Approximately 0.2 g of a stoichiometric mixture of chromium powder (99.97\%), red phosphorus pieces (99.999\%), and sulfur pieces (99.999\%), along with 0.01~g of iodine, were sealed in an evacuated quartz ampule ($\sim$12 cm long, $\sim$16 mm diameter). The ampule was placed in a two-zone furnace and subjected to a three-stage heating process. Initially, both zones were heated to 600~$^\circ$C over 58 hours. In the reaction stage, the deposition zone was raised to 680~$^\circ$C over 3 hours and maintained for 24 hours, while the charge zone remained at 600~$^\circ$C. During the transport stage, the temperature gradient was reversed, with the charge zone heated to 680~$^\circ$C and the deposition zone cooled to 600 $^\circ$C for 192 hours. The furnace was then allowed to cool naturally. This nucleation-controlled process resulted in the formation of \CrPS{} crystals several millimeters in length in the deposition zone. Bulk crystal characterization by x-ray diffraction and energy-dispersive x-ray spectroscopy are presented in \red{Supplementary Sec. 2}.

\CrPS{} flakes were exfoliated onto an O$_2$-plasma-cleaned Si/SiO$_2$ wafer inside an argon-filled glovebox on a hotplate at 80~$^\circ$C. Few-layer regions were identified by optical contrast and encapsulated by a thin sheet of hBN ($\sim$10 nm thick) from a polydimethylsiloxane stamp prior to removal from the glovebox. Scanning NV magnetometry measurements were performed using diamond probe tips (QZabre) inside a closed-cycle cryostat (attoDRY 2200) at a base insert temperature below 2 K. Stray field images were generally obtained at an optical power of 15~$\mu$W, or as low as 5 $\mu$W (e.g., odd-layer domain wall in Fig. \ref{fig:3}f,g) to minimize thermal disturbance to domain structures. The maximum accessible field for NV measurements was limited by our microwave hardware to 6 GHz or 0.3 T along the NV axis. Magnetic field ramps were performed at a ramp rate of 1 mT per second along any single axis with no deliberate pause at the maximum of the ramp. For fields along $B_{ext}^{NV}$ requiring adjustments along two axes, the field along the $z$-axis ($x$-axis) was ramped first when increasing (decreasing) $|B_{ext}^{NV}|$. Details of the experimental setup and measurement sequences are presented in \red{Supplementary Sec. 3}.

RMCD measurements were conducted at 2 K using an OptiCool cryostat with an out-of-plane magnetic field. The probe source was a He-Ne laser (632.8 nm) with its intensity modulated by a mechanical chopper (at 557 Hz) and polarization modulated by a photoelastic modulator (PEM, at 50.2 kHz). Laser power was maintained below 1~$\mu$W to prevent sample heating. A 50:50 beam splitter directed the back-reflected light to an avalanche photodiode detector. Two lock-in amplifiers isolated the reflection signals associated with the chopper and PEM frequencies. The RMCD signal was calculated as the ratio of these signals, multiplied by $\sqrt{3}$, and expressed as a percentage of total intensity. Each data set represents an average of at least 100 magnetic field sweeps to ensure reliability.

%\vspace{1em}
\subsection{Magnetic simulations}
Continuum micromagnetic simulations were performed using Ubermag \cite{Beg2022} with OOMMF as the computational engine \cite{Donahue1999}. Demagnetization fields were checked to not appreciably impact the domain wall trajectories and thus were not included in the final simulations presented. Additionally, atomistic spin dynamics simulations have been carried out based on the VAMPIRE software package \cite{Evans2014a}. The atomistic computational process consists of the following concatenated stages: (i) a zero field cooling from above critical to low non-zero temperatures, (ii) thermodynamic equilibrium at the target final temperature, and (iii) dynamic evolution in the presence of a magnetic field. On the one hand, processes (i) and (iii) have been carried out within the framework of the stochastic Landau-Lifshitz-Gilbert equation using a Heun numerical scheme, while for the equilibration period (ii) classical Monte Carlo calculations based on the Metropolis algorithm have been employed.

Although we assume equal magnitudes for the layer magnetizations in the atomistic simulations, we find that antiphase domain walls in even layers can nevertheless be field-driven above a critical field if the even layers are connected to odd layers (\red{Supplementary Videos 1-4}). Odd layers in opposite states on the two sides of the domain wall introduce a dipolar field gradient, as well as unequal exchange interactions on the two layer sublattices of the even layer (e.g., 5L next to 6L). In the presence of these asymmetries, an external field induces a net even-layer magnetization with a generally asymmetric spatial profile in the vicinity of the domain wall. This induced magnetization couples to inhomogeneous dipolar fields, including from the wall itself, to propel a directional propagation of the domain wall. Additional discussion of the micromagnetic and atomistic simulations are presented in \red{Supplementary Secs. 10 and 11}.

\subsection{Model for lateral exchange bias}
We employ a Stoner-Wohlfarth macrospin model, applicable to uniform rotations of single domains, to calculate the upper coercive field $B_{c2}$ for an odd-layer region in \CrPS{} surrounded by even-layer regions. The magnetization of the odd-layer region is parametrized by $\theta$ describing the angle between the magnetization of a particular layer and the uniaxial anisotropy easy axis. Antiparallel alignment between the moments in consecutive layers is assumed. The magnetization of the same layer in the connected even-layer (AFM) region is assumed to be fixed at $\theta = 0$, such that $B_{c2}$ references the $\theta = 0 \rightarrow \pi$ transition in the odd layer. The total magnetic energy of the odd layer is then given by 

%The direction of the magnetization for a single layer of the the odd-layer region is described by the single angle $\theta$ between the magnetization vector and the uniaxial anisotropy easy axis, while the magnetization of the AFM even-layer regions is assumed to be fixed at $\theta = 0$. Therefore, the total magnetic energy of the odd-layer region is given by 
\begin{equation}\label{eq:1}
\begin{split}
	E(\theta,\phi) = -\kappa \cos^2(\theta) S t_{odd}- B_{ext} \sigma_1 \cos(\theta-\phi) S\\ + \varepsilon_{DW}(1-\cos(\theta))C t_e / 2,
\end{split}
\end{equation}
where the three terms describe the anisotropy, Zeeman and interfacial domain wall energies, respectively. The external field $B_{ext}$ is applied at an angle $\phi$ relative to the easy axis. Here, $\kappa$ is the anisotropy energy per unit volume; $\varepsilon_{DW}$ is the domain wall energy density (per area); $S$, $t_{odd}$ and $\sigma_1$ are the area, thickness and magnetic moment per area of the odd layer; $C t_e$ denotes the surface area of the domain walls, given by the product between the perimeter and effective thickness (approximately the weighted average) of the surrounding even-layer regions. For a given portion of the interface, $t_e = \mathrm{min}(t_{odd}, t_{even})$, where $t_{even}$ is the thickness of the even layer.
 
For simplicity, we consider $\phi = 0$ ($B_{ext}$ parallel to the easy axis). $E(\theta,0)$ admits two possible local minima at $\theta = 0$ and $\theta = \pi$. We find that the local minimum at $\theta = 0$ is stable ($\partial^2E/\partial\theta^2 > 0$) for 
\begin{equation}\label{eq:2}
%\begin{split}
	B_{ext} > - 2 \kappa t_{odd}/\sigma_1 - \varepsilon_{DW} C t_e/ (2 \sigma_1 S),
%\end{split}
\end{equation}
where the first term describes the coercive field due to anisotropy and the second term represents an additional shift $B_E$ (exchange bias) due to the domain wall energy. The case for nonzero $\phi$ can be solved numerically. Both the width of the hysteresis curve and the additional exchange field $B_E$, induced by nonzero $\varepsilon_{DW}$, decrease as $\phi$ increases. Our experimental stray field linecuts (Supplementary Fig. S8) and atomistic simulations (Supplementary Fig. S21) find that domain walls in atomically thin \CrPS{} are predominantly Bloch type. This model is inapplicable to the $B_{c1}$ reversal of the odd layer ($\theta = \pi \rightarrow 0$) as it takes place by translation of the nucleated domain walls, rather than by uniform rotation.

\section{Data availability}
Source data for the main figures are provided with this paper. All data that support the findings of this study are available from the corresponding author upon request. 

\section{Code availability}
The codes that support the findings of this study are available upon request. Codes include scripts for data analysis and micromagnetic simulations.

\end{document}